\documentclass[12pt]{article}
\usepackage{graphicx}
\usepackage{dcolumn}
\usepackage{bm}
\usepackage{amsmath}   
\usepackage{graphics}
\usepackage{subfigure}
\begin{document}

\begin{center}

{\bf{\large Proton Mass Decomposition and Hadron Cosmological Constant}}

\vspace{0.6cm}


{\bf  Keh-Fei Liu}

\end{center}

\begin{center}
\begin{flushleft}
{\it
Department of Physics and Astronomy, University of Kentucky, Lexington, Kentucky 40506, USA
}
\end{flushleft}
\end{center}

\begin{abstract}

Lattice results on sigma terms and global analysis of parton momentum fractions are used to give the quark and glue fractions of the proton mass and rest energy. The mass decomposition in terms of the trace of the energy-momentum tensor is renormalization group invariant. The decomposition of the rest energy from the Hamiltonian and the gravitational form factors are scheme and scale dependent. The separation of the energy-momentum tensor into the traceless part which is composed of the quark and glue parton momentum fractions and the trace part has the minimum scheme dependence. 

We identify the glue part of the trace anomaly  $\langle H_{\beta}\rangle $ as the vacuum energy from the glue condensate in the vacuum. From the metric term of the gravitational form factors, which is the stress part of the stress-energy-momentum tensor, we find that the trace part of the rest energy, dominated by  $\langle H_{\beta}\rangle$, gives a {\it constant} restoring pressure that balances that from the traceless part of the Hamiltonian to confine the hadron, much like the cosmological constant Einstein introduced for a static universe. From a lattice calculation of $\langle H_{\beta}\rangle$ in the charmonium, we deduce the associated string tension which turns out to be in good agreement with that from a Cornell potential, which fits the charmonium spectrum.

\end{abstract}


.\section{Introduction} \label{intro}

    In this work, we consider the decompositions of the proton mass and rest energy into their respective quark and
glue components and will use lattice results and momentum factions from a global analysis to provide numerical results
for each component. Even though the mass and rest energy are equal in Einstein's equation i.e. $E_0 = m c^2$,
many properties associated with the mass and energy are not the same. For example, the mass is a Lorentz scalar
while the energy is the time component of the 4-momentum vector. In the example of $e^+ e^-$ annihilation to two photons
$e^+ e^- \longrightarrow \gamma\gamma$, the mass of the two photons from the rest energy is $2 m_e$, not the sum
of the two photon mass~\cite{Okun:1991nr,Okun:2000kf}. This shows that, while momentum and energy have additivity properties, mass does not. When there is mass there is energy, but not vice versa. For nonrelativistic particles, mass 
appears in Newtonian dynamics and gravitational interaction. In general relatively, however, the gravitational field is 
coupled to the energy-momentum tensor that has 10 components.~\footnote{The above discussion of mass and energy
can be found in L. Okun~\cite{Okun:1991nr,Okun:2000kf}.} 

      In the present work, we shall make a distinction between the mass and rest energy in the context of separating the
quark and glue components in the proton or other hadrons. As we shall see in the following text, the mass can be obtained
from the trace of the energy-momentum tensor (EMT). However, the fraction of each component, although renormalization group invariant, needs to be defined in the rest frame.  On the other hand, the decomposition of the proton rest energy, be it  through the Hamiltonian or the forward matrix elements of the gravitational form factors, is scheme and scale dependent. We shall evaluate the matrix elements associated with the operators to examine the physical meaning of them and; moreover, check if they are accessible experimentally and/or through lattice calculations. 

      The paper is organized as follows. In Sec.~\ref{trace}, we will discuss the proton mass decomposition via the trace of the energy-momentum tensor and use lattice QCD calculation of the sigma terms for quarks with different flavors to find
the fractional contribution for each flavor as well as the trace anomaly. In Secs.~\ref{Hamiltonian} and \ref{gravitational}, we will describe the decomposition of the rest energy in terms of the Hamiltonian and the forward matrix elements of the gravitational form factors. They are related to the quark and glue momentum fractions of the parton distribution functions which are obtained from the global analysis of experiments.  As such, they are scale dependent. There is quite a bit of interest lately in the issue on how to understand the physical meaning of each component in various ways of apportioning the proton mass. 
The first work was done by X. Ji who has decomposed the hadron rest energy from the Hamiltonian in terms of the quark sigma terms, the quark and glue energies and the anomaly~\cite{Ji:1994av,Ji:1995sv}. Lattice calculations based on this decomposition have been carried out for the mesons~\cite{Yang:2014xsa} and the nucleon~\cite{Yang:2018nqn}. Also, combined lattice and experimental results on the momentum fractions have been used
to evaluate the quark and glue components of the nucleon~\cite{Gao:2015aax}. Examining
the rest energy from the gravitational form factors, C. Lorc\'{e} has interpreted the quark and glue parts
in terms of the internal energies and pressure-volume work~ \cite{Lorce:2017xzd,Lorce:2018egm}. Y. Hatta, A. Rajan and K. Tanaka have explored the renormalization group property of the form factor $\bar{C}$ and performed perturbative calculation of the quark and glue parts of the trace of the energy-momentum tensor~\cite{Hatta:2018sqd}. The scheme dependence of the quark and glue
parts of the mass and rest energy has been further examined by A.~Metz, B.~Pasquini and S.~Rodini~\cite{Metz:2020vxd}.
Each of the above-mentioned work has some different expressions and interpretations in the decomposition which we will attempt to address and clarify. We shall discuss the significance of each term in the gravitational form factors in terms of thermodynamics. 
In Sec.~\ref{cosmological_constant}, we discuss the role of the scalar trace and tensor traceless parts of the rest energy 
in the metric term ($\eta^{\mu\nu})$ of the gravitational form factor which is the stress in the energy-momentum tensor. Comparing them in the energy and pressure equations, we find that the scalar part of the energy, dominated by
the glue part of the trace anomaly $\langle H_{\beta}\rangle$, has an energy density which is a constant so that it gives rise to
a constant restoring pressure to balance those from the quark and gluons and, thus, confines the proton. This shows that it is
the hadron cosmological constant. As a support of this idea, we show that the string tension deduced from a lattice calculation of $\langle H_{\beta}\rangle$ in the charmonium agrees well with that from a Cornell potential for the charmonium spectrum calculation. A summary is presented in Sec.~\ref{summary}.

\section{Mass Decomposition} \label{trace}

     The decomposition of proton mass and rest energy are all based on the energy momentum tensor (EMT).           
     The Belinfante form of the EMT is a symmetric rank two tensor 
\begin{equation}  \label{EMT}
T^{\mu\nu} = T_q^{\mu\nu} +  T_g^{\mu\nu},
\end{equation}
The bare operators are
\begin{eqnarray}  \label{EMT_bare}
T_q^{\mu\nu} &=& \frac{i}{4} \sum_f \bar{\psi}_f \gamma^{\{\mu}\!\stackrel{\leftrightarrow}{D}\!{}^{\nu\}}\psi_f
= \frac{i}{4} \sum_f \bar{\psi}_f (\stackrel{\rightarrow}{D}\!{}^{\mu}\gamma^{\nu} + \stackrel{\rightarrow}{D}\!{}^{\nu}\gamma^{\mu} - \stackrel{\leftarrow}{D}\!{}^{\mu}\gamma^{\nu} - \stackrel{\leftarrow}{D}\!{}^{\nu}\gamma^{\mu})\psi_f \\
T_g^{\mu\nu} &=& - G^{\mu\alpha} G^{\nu}_{\,\,\alpha} + \frac{1}{4} g^{\mu\nu} G^{\alpha\beta} G_{\alpha\beta}. \label{EMT_glue}
\end{eqnarray}
and
\begin{equation}
\stackrel{\leftrightarrow}{D}\!{}^{\mu} = \stackrel{\rightarrow}{\partial}\!{}^{\mu} - \stackrel{\leftarrow}{\partial}\!{}^{\mu}
-2 i g A_{a}^{\mu}\, T_a,
\end{equation}
where $T_a$ is the $SU(3)$ color matrix.

It is natural to consider the mass decomposition in terms of the trace of the EMT, since the matrix element of the trace gives
the nucleon mass and is frame independent
\begin{equation}
\langle P|T^{\mu}_{\mu}|P\rangle = 2 M^2.
\end{equation} 
On the other hand, the forward matrix element of the EMT component is 
\begin{equation}
\langle P|T^{\mu\nu}|P\rangle = 2 P^{\mu}P^{\nu},
\end{equation}
One can consider the rest energy from the $T^{00}$ component.

As far as the EMT trace is concerned, the classical trace is zero when the quark mass is neglected. However, this
conformal symmetry is broken by a trace anomaly in QCD due to quantum corrections~\cite{Chanowitz:1972vd,Crewther:1972kn,Chanowitz:1972da,Collins:1976yq}, 
\begin{equation}  \label{Trace_anomaly}
T_{\mu}^{\mu} =  \sum_f m_f (1+ \gamma_m (g)) \bar{\psi}_f \psi_f
 + \frac{\beta(g)}{2g} G^{\alpha\beta} G_{\alpha\beta},
 \end{equation}
where $\beta(g)$ is the $\beta$ function and $\gamma_m$ is the mass anomalous dimension.
In dimensional regularization, the quark condensate comes from the quark part  $T_{q \mu}^{\mu}$ in Eq.~(\ref{EMT_bare}) 
and the anomaly terms with $\gamma_m$ and $\beta$ are from $T_{g\, \mu}^{\mu}$, i.e.
\begin{eqnarray}  \label{trace_q,g}
T_{q\,\, \mu}^{\mu} &=&  \sum_f m_f \bar{\psi}_f \psi_f,  \label{trace_q} \\
T_{a\,\, \mu}^{\mu} &=&\sum_f m_f \gamma_m (g) \bar{\psi}_f \psi_f +\frac{\beta(g)}{2g} G^{\alpha\beta} G_{\alpha\beta} 
\label{trace_g}
\end{eqnarray}
It is pointed out that the above separation into quark and glue parts of the trace is scheme dependent and they
can mix under renromalization giving rise to scheme dependence~\cite{Hatta:2018sqd,Rodini:2020pis,Metz:2020vxd}.
However, since the EMT is conserved, i.e.
\begin{equation} \label{EMT_conservation}
\partial_{\nu} T^{\mu \nu}=  0 
\end{equation}
the renormalized $(T^{\mu\nu})_R$ is the same as the original one in Eqs.~(\ref{EMT}-\ref{EMT_glue}).
\begin{equation}
(T^{\mu\nu})_R=  T^{\mu\nu}.
\end{equation}
Therefore, the renormalized matrix elements of the trace are represented as
\begin{equation}  \label{Mass_total}
\langle (T^{\mu}_{\mu})_R \rangle = \sum_f \langle P| (m_f \bar{\psi}_f \psi_f)_R|P\rangle + \langle P|\frac{\beta(g_R)}{2g_R} (G^{\alpha\beta} G_{\alpha\beta})_R +\gamma_m(g_R) \sum_f (m_f\bar{\psi}_f\psi_f)_R|P\rangle = 2 M^2,
\end{equation}
where $\langle ... \rangle = \langle P|...|P\rangle$.
%
%
As long as one does not take the renormalized $T_{q\,\, \mu}^{\mu} $ and $T_{a\,\, \mu}^{\mu}$ in Eqs.~(\ref{trace_q}) and
(\ref{trace_g}) as separate quark and glue contributions, but considers the renormalized $\langle (T^{\mu}_{\mu})_R \rangle$ as a whole, the expression in Eq.~(\ref{Mass_total}) is scheme independent and renormalization group invariant. However, the matrix element is proportional to $M^2$, not $M$ itself. This raised the question and ambiguity about its proper normalization~\cite{Lorce:2018egm,Hatta:2018sqd,Metz:2020vxd,Ji:2021mtz,Roberts:2016vyn}. The usual definition of an expectation value $\frac{\langle \int d^3 \vec{x}\,T^{\mu}_{\mu}(r)\rangle}{\langle P|P\rangle} = \frac{M^2}{P^0}$ depends on the frame, since $\langle P|P\rangle = (2\pi)^3 2 P^0 \delta^3(0)$.  One exception is when all the components are in the rest frame. In this case, $\frac{\langle \int d^3 \vec{x}\,T^{\mu}_{\mu}(r)\rangle}{\langle P|P\rangle}|_{\vec{P} = 0}= M$. In the moving frame, one should take the integral $\int d^3 \vec{x}\, \gamma$ to be over the proper volume so that $\frac{\langle \int d^3  \vec{x}\, \gamma T^{\mu}_{\mu}(r)\rangle}{\langle P|P\rangle}= M$, where $\gamma = \frac{1}{\sqrt{1 - v^2/c^2}}$~\footnote{Thanks to C. Lorce for bringing about this point.}. This shows that mass is frame independent.

In this work, we shall define the fractional contribution of each term in Eq.~(\ref{Mass_total})
\begin{eqnarray}  \label{f_mass}
f_f^N &=& \frac{\langle P| \int d^3 \vec{x} \, m_f \bar{\psi}_f \psi_f|P\rangle}{M \langle P|P\rangle}|_{\vec{p} = 0}  \label{f_f}  \\
f_{\rm{a}}^N &=& \frac{\langle P| \int d^3 \vec{x}\, [\frac{\beta(g_R)}{2g_R} (G^{\alpha\beta} G_{\alpha\beta})_R +\gamma_m(g_R) \sum_f (m_f\bar{\psi}_f\psi_f)_R]|P\rangle}{M \langle P|P\rangle}|_{\vec{p} = 0} \label{f_a}
\end{eqnarray}
where the quark mass times the quark condensate is know as the sigma term and the fraction $f_f^N$ is thus the mass fraction of
the sigma term
\begin{eqnarray}
\sigma_{\pi N} &=& \frac{m_u+m_d}{2} \langle P|\bar{u}u + \bar{d}d|P\rangle_{\vec{P} =0}/2M, \hspace{1.5cm} f_{\pi N}^N = f_u^N + f_d^N =\frac{\sigma_{\pi N}}{M} , \\
\sigma_{f = (s, c, b, t)} &=& m_f \langle P|\bar{\psi}_f \psi_f|P\rangle_{\vec{P}=0}/2M, \hspace{2.21cm} f_{f=(s,c,b,t)}^N = \frac{\sigma_{f=(s,c,b,t)}}{M}.
\end{eqnarray}

\begin{figure}[htbp]     
\centering
\subfigure[]
{\includegraphics[width=0.445\hsize]{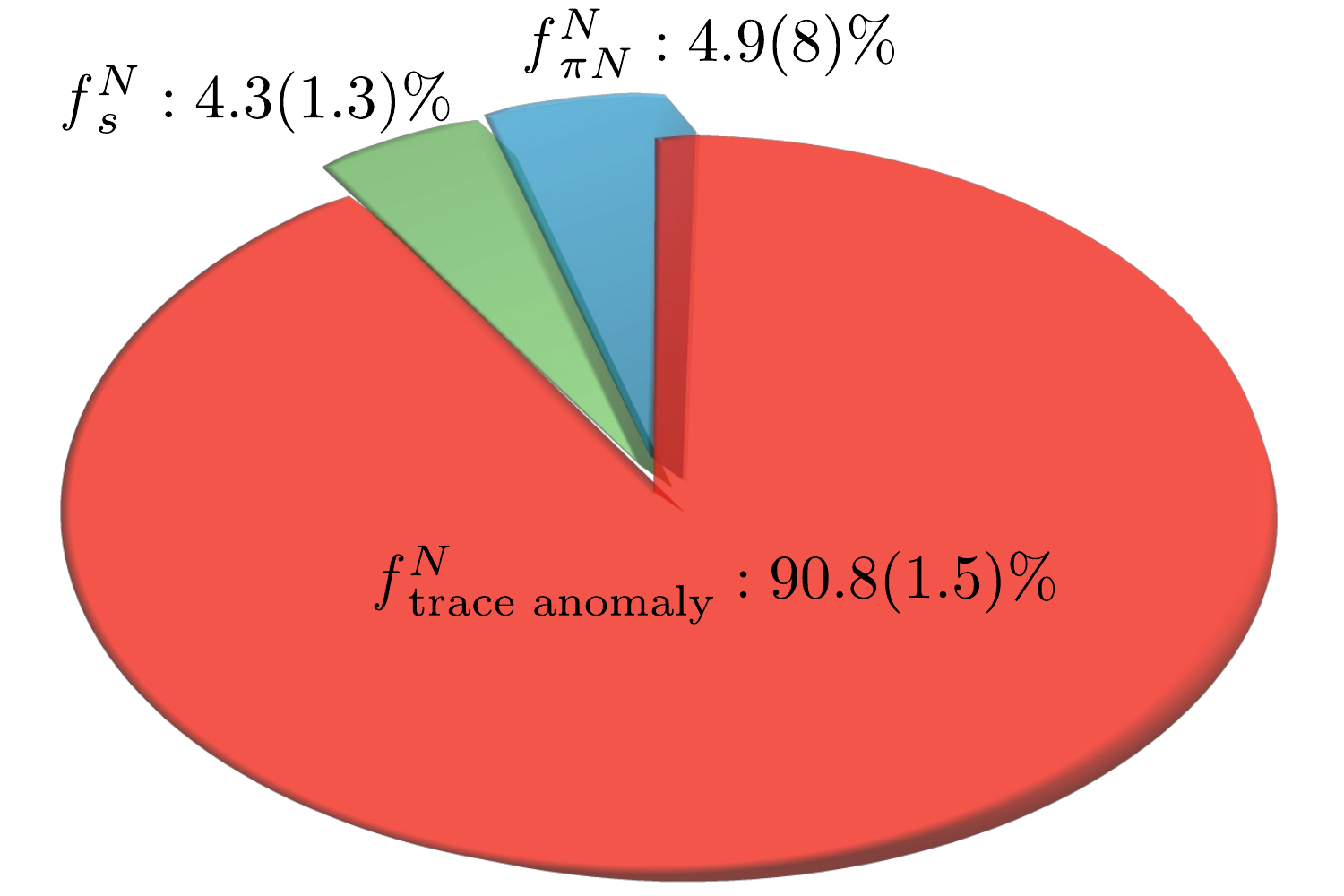}
  \label{mass_3f}}
\subfigure[]
{\includegraphics[width=0.445\hsize]{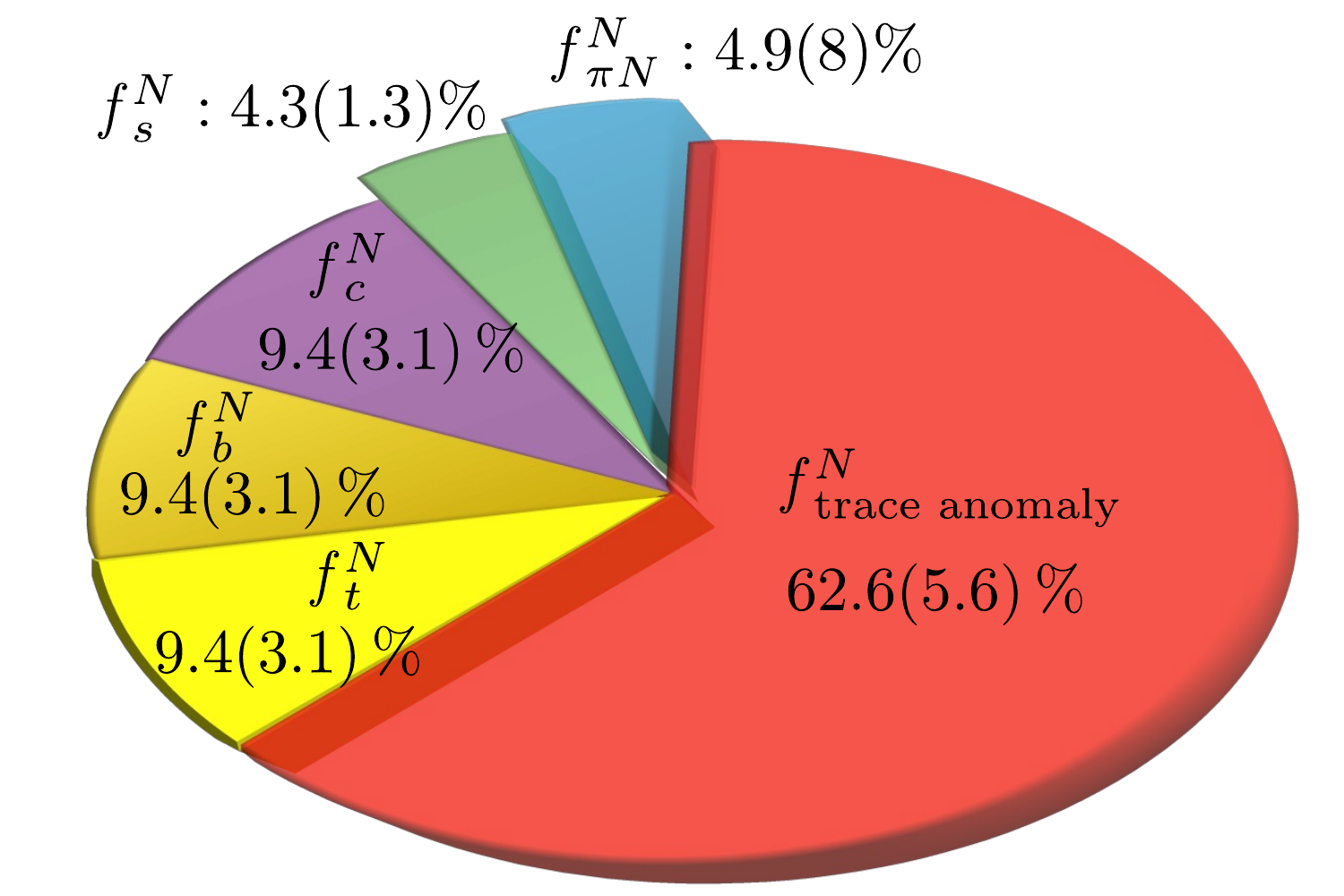}
  \label{mass_6f}}
\caption{Proton mass decomposition in terms of quark sigma terms of different flavors and the trace anomaly. They are plotted as the percentage fractions of the proton mass. The sigma terms are obtained from lattice 
calculations~\cite{Yang:2015uis,Gong:2013vja}. (a) is for the 2+1 flavor case at the hadron scale $\sim 1$ GeV and (b) for the case including the charm, bottom and top sigma terms.}
  \end{figure}

The nucleon sigma terms have been calculated on the lattice and they are tabulated in FLAG~\cite{Aoki:2019cca} . We shall use the results from the overlap fermion~\cite{Neuberger:1997fp} that has exact chiral symmetry on the lattice \`{a} la Gingparg-Wilson relation~\cite{Ginsparg:1981bj}. In this calculation~\cite{Yang:2015uis}, several lattice ensembles with different lattice
spacings and quark masses with one lattice at physical pion mass are used to address the systematic errors in the continuum and infinite volume extrapolations at the physical pion point. For QCD at hadronic scale $\sim 1$ GeV, the heavy quarks are 
integrated out so that they are reflected in the $\beta$ function. In Table~\ref{tab:mass}, we give $f_{\pi N}^N$ and $f_s^N$ for the $(2+1)$-flavor case. The trace anomaly contribution $f_a^N$ is obtained from subtracting the $\sum_f f_f^N$ from unity.
$f_a^N$ can in principle be calculated on the lattice which involves more elaborate renormalization~\cite{Panagopoulos:2020qcn} and will have larger errors than that obtained here from the mass sum rule.
As we see from Table~\ref{tab:mass} and Fig.~\ref{mass_3f} that the sigma terms contribute only $\sim$ 9\% to the nucleon mass. The rest comes from the trace anomaly. 

When one considers QCD as part of the standard model at the weak scale, the Higgs and heavy quarks (c, b, t) are all involved as relevant operators and can appear in external states. This is relevant to high energy
processes such as DIS and high energy hadron collisions. If the dark matter, such as neutrilino, couples to the Higgs, then the heavy quark sigma terms will contribute to the scattering cross section of the dark matter on nucleus~\cite{Hill:2014yxa}. It was
pointed out~\cite{Shifman:1978zn} that, to the leading order in the heavy quark expansion, 
the matrix element $m_Q \langle P | \bar{Q} Q | P \rangle$ is related to the glue condensate in the nucleon,
\begin{equation}  \label{gc}
\sigma_Q \equiv m_Q \langle P | \bar{Q} Q | P \rangle_{\stackrel{\longrightarrow}{m_Q \rightarrow \infty}} - \frac{1}{3}(\frac{\alpha_s}{4 \pi})\langle  P | G^2 |P \rangle.
\end{equation}
This is precisely the $n_f$ term in the leading $\alpha_s/4\pi$ expansion of $\frac{\beta(g)}{2g} = - \frac{\beta_0}{2} (\frac{\alpha_s}{4\pi}) - \frac{\beta_1}{2} (\frac{\alpha_s}{4\pi}) ^2 + ...$~\cite{Chetyrkin:1997dh,Vermaseren:1997fq}, where $\beta_0 = (11 - \frac{2}{3} n_f)$, except with
a negative sign. This shows that for $n_f$ heavy enough quarks, the introduction of their sigma terms can trade with
the $n_f$ term in $\beta_0$, the leading term in $\beta(g)/2g$. To study the quark mass dependence, a lattice
calculation with the overlap fermion has been carried out~\cite{Gong:2013vja}. It is found that the sigma terms for quark masses heavier that  $\sim 1/2$ of the charm mass are the same within errors. We take this finding to mean that the sigma terms for the charm, beauty and top quarks are the same. For the charm, it is found~\cite{Gong:2013vja} that $f_c^N = 0.094(31)$, which is taken to be the same for $f_b^N$ and $f_t^N$. The sigma terms for the heavy quarks are listed in Table~\ref{tab:mass} and the
pie chart for the quark sigma terms for all six flavors and the corresponding trace anomaly term is given in Fig.~\ref{mass_6f}.
Even though the heavy quark sigma terms are listed in Table~\ref{tab:mass} and plotted in Fig.~\ref{mass_6f}, it should not be
misconstrued to imply that they are the total heavy quark contributions to the proton mass. As explained above, the heavy quarks
also contribute negatively through the $\beta$ function with $n_f$ flavor [cf. Eq.~(\ref{gc})] so that the net contribution of
a heavy quark with mass $M_H$ is $\mathcal{O} (1/m_H)$, in accordance with the decoupling theorem~\cite{Appelquist:1974tg,Kaplan:1988ku}.

 \begin{table}[h]
  \centering
  \caption{Decomposition of proton mass in terms of the quark sigma terms of different flavors and the trace anomaly. They are 
 given as the percentage fractions of the proton mass. The sigma terms are obtained from lattice calculations~\cite{Yang:2015uis,Gong:2013vja}.}
\bigskip
  \begin{tabular} {|c|ccccc|cc|}
  \hline
 $n_f$ & $ f_{\pi N}^N$ & $ f_s^N$ & $ f_c^N$ &  $f_b^N$ & $f_t^N$ & $f_{q \,\rm{total}}^N$ & $f_{\rm{a}}^N$ \\
 \hline
     2+1  & 4.9(8)  &  4.3(1.3) &  ...  & ... & ... & 9.2(1.5) & 90.8(1.5)  \\
    \hline
    2+1+1+1+1  & 4.9(8)  & 4.3(1.3) & 9.4(3.1) & 9.4(3.1) & 9.4(3.1) & 37.4 (5.6) & 62.6 (5.6)  \\
   \hline
      \end{tabular}   \label{tab:mass}
   \end{table}
%

\section{Rest Energy from Hamiltonian}  \label{Hamiltonian}

\begin{figure}[htbp]     
\centering
\subfigure[]
{\includegraphics[width=0.475\hsize]{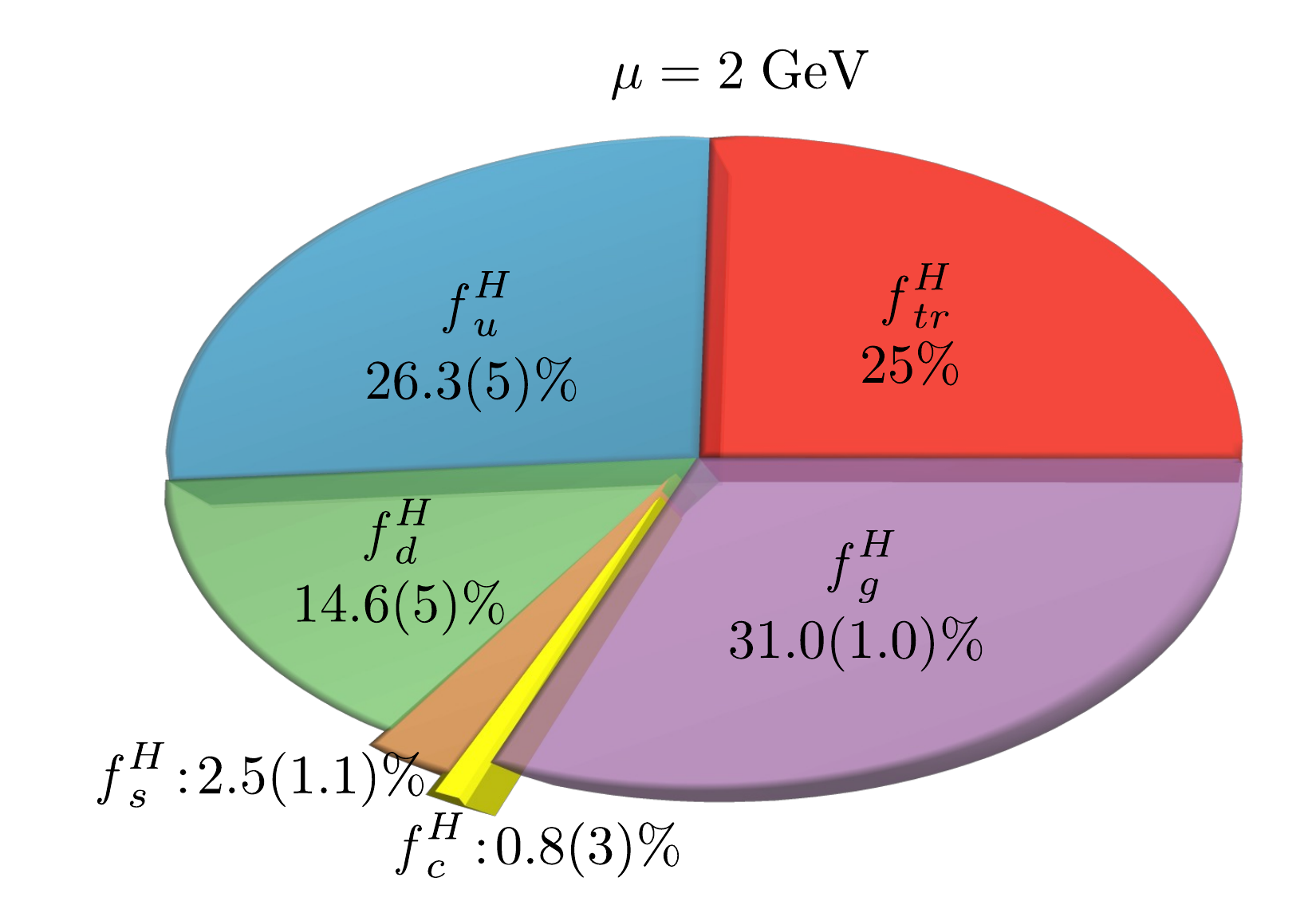}
  \label{h_2GeV}}
\subfigure[]
{\raisebox{0mm}{\includegraphics[width=0.475\hsize]{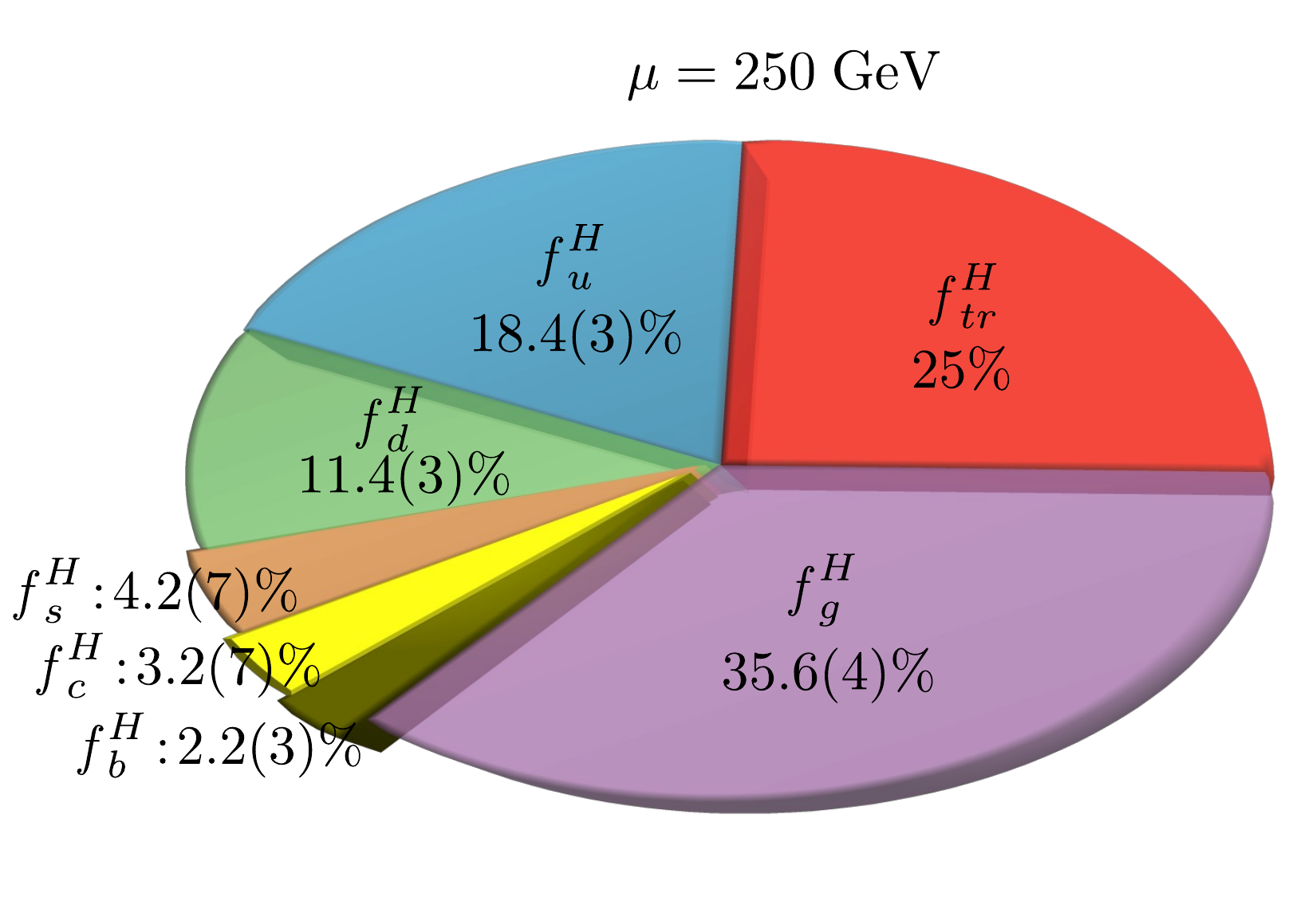}
  \label{h_250GeV}}}
\caption{Proton rest energy in terms of the quark mass-energy, the glue field energy and the trace of the
EMT which is a quarter of the proton mass. They are given as the percentage fractions of the proton mass. The traceless parts
of the EMT are related to the quark and glue momentum fractions on the light front and they are given from
the CT18 global analysis of DIS and Drell-Yan experiments. (a) is for the decomposition at $\mu = 2$ GeV (b) for $\mu = 250$ GeV.}
 \end{figure}

     We shall consider the rest energy from the $T^{00}$ component of the EMT matrix elements at rest. There are several ways to examine the decomposition into quark and glue contributions. The decomposition of the proton rest energy with the Hamiltonian was first considered by X. Ji~~\cite{Ji:1994av,Ji:1995sv}. Since the EMT in the Belinfante form is a symmetric
rank two tensor, they can be separated into the traceless and trace parts. They are in different representations of the Lorentz
group. The traceless and trace parts are in the irreducible (1,1) and (0,0) representations respectively. Hence, they do not mix. The renormalized EMT is  separated as
\begin{equation} \label{trace_separation}
(T^{\mu\nu})_R = (\overline{T}^{\mu\nu})_R + (\hat{T}^{\mu\nu})_R,
\end{equation}
where $ (\hat{T}^{\mu\nu})_R = \frac{1}{4} \eta^{\mu\nu} T_{\rho}^{\rho}$. So far, this separation is
scale and scheme independent. $\bar{T}^{\mu\nu}$ can be further split into the quark and glue parts. In this case,
the Hamiltonian, being the spatial integral of $T^{00}$, i.e. $ H = \int d^3\vec{x}\, T^{00}(x)$, can be written as
\begin{equation}  \label{H_3-term}
H = H_q (\mu)+ H_g (\mu)+ H_{tr},
\end{equation}
where 
 \begin{eqnarray}
 H_q (\mu)&=& \int d^3\vec{x}\, (\frac{i}{4} \sum_f \bar{\psi}_f \gamma^{\{0}\!\stackrel{\leftrightarrow}{D}\!{}^{0\}}\psi_f 
 - \frac{1}{4} T_{q\, \mu}^{\mu})_R, \\
 H_g (\mu)&=& \int d^3\vec{x}\, \frac{1}{2} (B^2 + E^2)_R,  \label{H_g} \\
 H_{tr} &=& \int d^3\vec{x}\,  \frac{1}{4} (T_{\mu}^{\mu})_R.
  \end{eqnarray}  
Here, the subscript $R$ indicates it is renormalized and mixed if needed. For example, the separation of $H_q$ and $H_g$ from the traceless $\overline{T}^{\mu\nu}$ entails operator mixing besides renormalization and the scale dependence is introduced. These nucleon matrix elements can be extracted from experiments and lattice calculations at the scale $\mu$. 
$H_q (\mu)$ is the quark mass and energy operator and $H_g(\mu)$ corresponds to the glue field energy operator.
Their matrix elements are related to the quark and glue momentum fractions
\begin{eqnarray}
 \langle H_{q} (\mu)\rangle  &=& \frac{3}{4} \sum_f \langle x\rangle_f (\mu) M, \\
 \langle H_g (\mu) \rangle &=& \frac{3}{4}  \langle x\rangle_g (\mu) M, \\
 \langle H_{tr} \rangle&=& \frac{1}{4} M.
  \end{eqnarray}   
where $\langle x\rangle_f(\mu)$ and $\langle x\rangle_g(\mu)$ are the momentum fractions for the quark with flavor $f$ and the glue from DIS or Drell-Yan experiments at the scale $\mu$ and they satisfy momentum conservation, i.e. $\sum_f \langle x\rangle_f (\mu) + \langle x\rangle_g (\mu) = 1$. This decomposition gives a sum rule for the nucleon rest energy 
\begin{equation}   \label{3-term_energy}
E_0 = M =  \langle H_{q} (\mu)\rangle +  \langle H_g (\mu) \rangle +  \langle H_{tr} \rangle,
\end{equation}
Here, we have defined $\langle H_{...}\rangle = \langle P|H_{...}|P\rangle/\langle P|P\rangle$ at $\vec{P} = 0$.
This is the simplest and well-defined separation of the rest energy from the Hamiltonian---each term is related to
experimental and/or lattice observables, with the necessary scale dependence in the $\overline{\rm{MS}}$ scheme and there is no further scheme dependence for the trace term. We shall use $\langle x\rangle_f(\mu)$ and $\langle x\rangle_g(\mu)$ from CT18 global
analysis of the PDFs~\cite{Hou:2019efy} for the rest energy decomposition. They are listed in Table~\ref{tab:x} for $\mu = 2$ and 250 GeV. As can be seen in Table~\ref{tab:x}, the momentum fractions of the $u$ and $d$ partons, which are mostly due to the valence contributions,
are shifted to those of the sea-quark and glue partons as the scale $\mu$ increases from 2 to 250 GeV, the electroweak scale. We shall define the fractions
\begin{eqnarray}
f_f^H &=&\langle H_{q}\rangle/M = \frac{3}{4} \langle x\rangle_f(\mu), \label{f_qf}\\
f_g^H  &=&\langle H_{g}\rangle/M = \frac{3}{4} \langle x\rangle_g (\mu), \label{f_g}\\
f_{tr}^H   &=&   \langle H_{tr} \rangle/M = \frac{1}{4}
\end{eqnarray}
and plot them in Fig.~\ref{h_2GeV} for $\mu = 2$ GeV and Fig.~\ref{h_250GeV} for $\mu = 250$ GeV. We note
that $f_{tr}^H$ can be further decomposed as in Eqs. (\ref{f_f}) and (\ref{f_a}). It is shown 
that this separation can have a Lorentz-invariant interpretation~\cite{Lorce:2017xzd,Cosyn:2019aio,Ji:2021mtz}.

\begin{center}
   \begin{table}[h]
  \centering
  \renewcommand{\arraystretch}{1.4}
  \caption{The quark and glue momentum fractions $\langle x\rangle_f(\mu)$ and $\langle x\rangle_g(\mu)$ from CT18 global
analysis of the PDFs~\cite{Hou:2019efy} are tabulated in the $\overline{MS}$ scheme for $\mu = 2$ GeV in the second row and $\mu = 250$ GeV in the third row~\cite{T.J.Hou}. They are given as percentages of the total proton momentum.}
\vspace{1cm}
  \begin{tabular} {|c|cccccc|cc|}
  \hline
 $$ & $ u$ & $ d$ & $ s$ & $ c$ &  $ b$ & $t$ & $ \rm{q_{total}}$ &  glue \\
 \hline
 $ \langle x\rangle_{f,g}$ (2 GeV)  & 35.0(7)  & 19.4(7) & 3.3(1.5)   & 1.1(4)   & 0 & 0 & 58.7(1.5) & 41.3(1.4)  \\
    \hline
 $ \langle x\rangle_{f,g}$(250 GeV)  & 24.5(4)  & 15.2(4) & 5.6 (9)  & 4.3(9)  & 2.9(4) & 0 & 52.5(1.3) & 47.5(5)  \\
    \hline
 \end{tabular}   \label{tab:x}
 \end{table}
  \end{center}
There are suggestions to separate $H_{tr}$ into the sigma terms and the trace 
anomaly~\cite{Ji:1994av,Ji:1995sv,Metz:2020vxd} so that the Hamiltonian is
\begin{equation}  \label{4-term_trace}
H = H_q (\mu)+ H_g (\mu)+\frac{1}{4} H_m + \frac{1}{4}  H_a
\end{equation}
where 
\begin{eqnarray}
H_m &=& \int d^3\,\vec{x}\, \sum_f m_f \bar{\psi}_f \psi_f,  \label{H_m} \\
H_a &=&  \int d^3\,\vec{x}\, \big[\sum_f m_f \gamma_m (g) \bar{\psi}_f \psi_f +\frac{\beta(g)}{2g} G^{\alpha\beta} G_{\alpha\beta}
\big]
\end{eqnarray}
However, it was pointed out that the separation of $H_{tr}$ introduces
a scheme dependence~\cite{Hatta:2018sqd,Rodini:2020pis,Metz:2020vxd} and the separation suggested here amounts to taking a specific D2 scheme~\cite{Metz:2020vxd}. Nevertheless, this is a physically motivated separation in the sense that
$H_m$ and $H_a$ are each scale invariant and they can be determined from lattice calculations as we alluded to in
Sec.~\ref{trace}.
 It has also been suggested~\cite{Ji:1994av,Ji:1995sv} to further separate the $H_q + \frac{1}{4} H_m$ into
 $H_m$ and $H_E$  by applying the equation of motion,
 \begin{equation}   \label{4-term}
H = H_m + H_E (\mu)+ H_g (\mu) + \frac{1}{4}  H_a.
\end{equation}
The classical form of $H_E(\mu)$ is $\psi^{\dagger} i \vec{\alpha} \cdot \vec{D}\psi$ and can be interpreted as
the quark kinetic and potential energy. The renormalized matrix element of $H_E$ has mixings with
both $H_g$ and $H_m$. The mixing of the traceless matrix elements is 
\bigskip
\begin{equation}  
\begin{pmatrix} \langle H_q \rangle_R (\mu)\\ \langle H_g\rangle_R (\mu)\end{pmatrix}
= \begin{pmatrix} Z_{qq} (\mu)& Z_{qg} (\mu)\\ Z_{gq} (\mu) & Z_{gg} (\mu) \end{pmatrix}
\begin{pmatrix} \langle H_q \rangle (\mu_r) \\ \langle H_g\rangle (\mu_r)\end{pmatrix}
\end{equation}
where $Z_{qq} (\mu)= 1 - Z_{gq}(\mu)$ and $Z_{gg} (\mu)= 1 - Z_{qg}(\mu)$ to preserve momentum conservation and
$\mu_r$ is a reference scale for the unmixed matrix element, such as the lattice spacing in lattice calculations.
The renormalization group equation for the evolution of the moments, to first order, is
\bigskip
\begin{equation}
\frac{\partial}{ \ln\mu} \begin{pmatrix} \langle x \rangle_q (\mu)\\ \langle x\rangle_g (\mu)\end{pmatrix}
= \frac{\alpha_s}{4\pi}
\begin{pmatrix} - \frac{16}{3}C_F & \frac{4}{3} n_f\\ \frac{16}{3}C_F & \frac{4}{3} n_f \end{pmatrix}
\begin{pmatrix} \langle x \rangle_q (\mu)\\ \langle x\rangle_g (\mu) \end{pmatrix}
\end{equation}
where $C_F = 4/3$.
The renormalized and mixed $(H_E)_R (\mu)$ is then
\begin{equation}  \label{H_E}
 (H_E)_R (\mu) = \sum_f (\psi_f^{\dagger} i \vec{\alpha}\cdot\vec{D}\psi_f)_R + 
 \big[-Z_{gq} H_m + \frac{4}{3} Z_{qg} H_g (\mu_r) \big].
 \end{equation} 
 The first term on the right-hand side is the self-renormalized operator and the terms in the square bracket are
 from mixing with $H_m$ and $H_g$.
 %
The matrix element of $H_E (\mu)$ in terms of $\langle x\rangle (\mu)$ and $\langle H_m\rangle$ turns out to be
  \begin{equation}  
 \langle H_E \rangle (\mu) = \frac{3}{4} (\langle x\rangle_q (\mu) - \langle H_m\rangle),
 \end{equation}     
 which is the same as in Refs.~\cite{Ji:1994av,Ji:1995sv} so that the rest energy sum rule
 \begin{equation}   \label{sum_rule_4_term}
 E_0 = M = \langle H_m\rangle +  \langle H_E (\mu)\rangle + \langle H_g (\mu)\rangle  + \frac{1}{4}  \langle H_a\rangle
 \end{equation}
 is satisfied. 
 
 In an effort to have a simplified decomposition with physical interpretation for each term, Metz, Pasquini and Rodini~\cite{Metz:2020vxd} have derived a decomposition based on dimensional regularization in the $\overline{\rm{MS}}$ scheme. It is composed of three terms
 \begin{equation}
 H = H_m' + H_E' + H_g'
 \end{equation}
 which represent the quark mass, quark energy and glue energy respectively. They have the same formal expressions for the corresponding operators as in Eqs.~(\ref{H_m}), (\ref{H_g}), and ({\ref{H_E}) for $H_m$, $H_g$  and $H_E$ (without the perturbative $Z$ terms), but without $H_a$ as compared to Eq.~(\ref{4-term}). The difference appears to be due to the definition of the operators. When the matrix elements are involved, 
 \begin{equation} 
 \langle H_g'\rangle = \frac{3}{4} \langle x\rangle_g + \frac{1}{4}\langle H_a \rangle,
\end{equation}
which is the sum of $\langle H_g\rangle$ and $\frac{1}{4} \langle H_a \rangle$. Thus, in the end, the same
sum rule is satisfied as in Eq.~(\ref{sum_rule_4_term}).
    
\section{Rest Energy from Gravitational Form Factors}  \label{gravitational}

    The rest energy has also been discussed in the context of the gravitational form factors. These are form factors for
the EMT and contain the following terms~\cite{Kobzarev:1962wt,Pagels:1966zza,Ji:1996ek,Ji:1996nm}
for the quarks and gluons 
\begin{eqnarray} \label{EMT_FF}
\langle P'| (T_{q, g}^{\mu\nu})_R(\mu)|P\rangle /2 M&=& \bar{u}(P')[T_{1_{q,g}}(q^2,\mu) \gamma^{(\mu} \bar{P}^{\nu)} +
T_{2_{q,g}}(q^2,\mu) \frac{\bar{P}^{(\mu} i \sigma^{\nu)\alpha} q_{\alpha}}{2M}   \nonumber \\
     &+&  D_{q,g}(q^2,\mu)\frac{q^{\mu}q^{\nu} - \eta^{\mu\nu}q^2}{M} + \bar{C}_{q,g}(q^2, \mu) M \eta^{\mu\nu} ] u(P)
\end{eqnarray}
where $T_{1_{q,g}} (0)= \langle x\rangle_{q,g} (\mu)$ is the momentum fraction and $T_{1_{q,g}} (0) + T_{2_{q,g}} (0) = 2J_{q,g} (\mu)$ the angular momentum fraction~\cite{Ji:1996ek,Ji:1996nm}. From the conservation of momentum and angular momentum, one has the following sum rules~\cite{Ji:1996ek,Ji:1996nm}
\begin{eqnarray}
&& T_{1_{q}} (0) + T_{1_{g}} (0) =\langle x\rangle_{q} (\mu) + \langle x\rangle_{g} (\mu) = 1, \nonumber \\
&& T_{1_{q}} (0) + T_{2_{1_q}} (0) + T_{1_{g}} (0) + T_{2_{1_g}} (0) = 2 J = 1.
\end{eqnarray} 
$T_1$ and $T_2$ are named after $F_1$ and $F_2$ of the electromagnetic form factors due to their analogy
to the form factor structure of the vector current, where the forward 
$F_1(0)$ is the charge and  forward $F_1(0) + F_2(0) = \mu_{p,n}$ the magnetic moments of the nucleon.
By making a connection to the stress tensor of the continuous medium, it is shown by Polyakov~\cite{Polyakov:2002yz}
and Polyakov and Schweitzer~\cite{Polyakov:2018zvc} that $D_{q,g}(q^2)$ is related to the internal force of the hadron and
encodes the shear forces and pressure distributions of the quarks and glue in the nucleon. 
The $D$ term, which is $D(0)$, has been emphasized~\cite{Polyakov:2018zvc} to be
the least known fundamental constant of the nucleon as compared to charge, magnetic moments, $g_A$, $g_P$, etc.
The pressure distribution of the quarks have been deduced from the experimentally measured $D_q(q^2)$ ~\cite{Burkert:2018bqq} and the pressure distributions for both the quarks and glue from  $D_{q,g}(q^2)$ are 
calculated on the lattice~\cite{Shanahan:2018nnv}.

$\bar{C}$ has the constraint in that the quark and glue parts cancel due to EMT conservation in Eq.~(\ref{EMT_conservation}), i.e.
\begin{equation}   \label{C-pressure}
\bar{C}_{\rm total}(0, \mu) = \bar{C}_q (0, \mu)+ \bar{C}_g (0, \mu)= 0.
\end{equation}
For the forward matrix elements at rest, Eq.~(\ref{EMT_FF}) becomes
\begin{eqnarray}
\langle P|(T_{q,g}^{00})_R(\mu)|P \rangle / 2M &=&  \langle x\rangle_{q,g} (\mu) M+ \bar{C}_{q,g}(0, \mu) M, \label{T00}\\
\langle P|(T_{q,g}^{ii})_R(\mu)|P \rangle / 2M &=&  - 3 \,\bar{C}_{q,g}(0, \mu) M \label{Tii}
\end{eqnarray}
%
%
From Eqs.~(\ref{T00}), (\ref{Tii}), (\ref{trace_q}), (\ref{trace_g})  and (\ref{Mass_total}), we obtain
\begin{eqnarray}   \label{bar-C}
\bar{C}_{q} (0, \mu) = \frac{1}{4} \sum_f (f_{f}^N- \langle x\rangle_{f}(\mu)) , \nonumber \\
\bar{C}_{g} (0, \mu) = \frac{1}{4}  (f_{a}^N- \langle x\rangle_{g}(\mu)) ,
\end{eqnarray}
Thus, Eq. (\ref{T00}) reproduces the mass sum rules in Eqs. (\ref{3-term_energy})  and (\ref{4-term_trace}) from the Hamiltonian without having to be separated into the trace and traceless parts explicitly. This presumably is due to the fact that
Lorentz symmetry has been implemented in the classification of the gravitational form factors in Eq.~(\ref{EMT_FF}).
We find that at $\mu = 2$ GeV, $\bar{C}_{u+d}(0), \bar{C}_s(0), \bar{C}_c(0) $ are -0.124(3), 0.003 (5), and -0.003(1), respectively for a total  $\bar{C}_{q} (0, \,\rm{2\, GeV}) = -0.124 (6)$. At $\mu = 250$ GeV, the corresponding $\bar{C}_{u+d}(0), \bar{C}_s(0), \bar{C}_c(0) \bar{C}_b (0), \bar{C}_t(0)$ are -0.087(2), -0.003(17), 0.013(8), 0.016(8) and 0.024(8)
 for a total $\bar{C}_{q} (0,\, \rm{250\, GeV}) = -0.038(14)$. We see that at the
hadronic scale $\bar{C}_{q} (0)$ is negative and its absolute value decreases as the scale increases. It is in agreement
with the estimate of $- 0.11$ at $\mu = 2$  GeV~\cite{Lorce:2017xzd}. However, it is different in sign from 
$\bar{C}_{q} (0) \sim 0.014$ estimated from the instanton vacuum~\cite{Polyakov:2018exb}. 

In view of the similarity between the structure  of the gravitational form factor in Eq.~(\ref{EMT_FF}) 
\begin{equation}
\langle P| (T_{q, g}^{\mu\nu})_R(\mu)|P\rangle /2 M V = T_{1_{q,g}} (0) P^{\mu}P^{\nu} + \eta^{\mu\nu} \bar{C} (0)
\end{equation}
and that of the stress tensor of the perfect fluid frequently used in general relativity 
\begin{equation}
T^{\mu\nu} = (\epsilon + p) u^{\mu} u^{\nu} - p\, \eta^{\mu\nu},
\end{equation}
where $u^{\mu} = P^{\mu}/M$, C. Lorc\'{e} identified the quark and glue parts of the EMT in Eq.~(\ref{EMT_FF}) as two fluids~\cite{Lorce:2017xzd,Lorce:2018egm} with
\begin{eqnarray}
\epsilon_{q,g} &\equiv& \big[T_{1_{q,g}} (0) + \bar{C}_{q,g} (0)\big] \frac{M}{V},  \\
p_{q,g} &\equiv& -\, \bar{C}_{q,g} (0) \frac{M}{V},
\end{eqnarray}
where $V$ is the proper volume. Therefore, the rest energy has been interpreted in terms of the following thermodynamic functions
\begin{eqnarray}   \label{U_qg}
U_q\!\! &= &\!\! \epsilon_q V=  \big[\langle x\rangle_q + \bar{C}_q (0)\big]M = \big[\frac{3}{4}\langle x\rangle_q + \frac{1}{4} \sum_f f_f^N\big]M,  \nonumber \\
U_g \!\!&= &\!\! \epsilon_g V =  \big[\langle x\rangle_g + \bar{C}_g (0)\big]M = \big[\frac{3}{4}\langle x\rangle_q 
+ \frac{1}{4}  f_ a^N\big]M. 
\end{eqnarray}
Here $U_q$  and $U_q$ are the internal energies for the quarks and glue and they are equal to the matrix elements of $H_q + \frac{1}{4} H_m$ and $H_g + \frac{1}{4}H_a$ in Eq.~(\ref{4-term_trace}), respectively. The work $W$ is
\begin{equation}  \label{work}
W_{q,g} = p_{q,g} V = - \,\bar{C}_{q,g} (0) M.
\end{equation}
Since the enthalpy is $H = U +  W$, one finds the enthalpies for quarks and glue are
\begin{equation}
H_{q,g} = U_{q,g} + W_{q,g} = \langle x\rangle_{q,g}.
\end{equation}
The rest energy and total work are the sum of the quark and glue internal energies and their works
\begin{eqnarray}
E_0 &=& U_q + U_g, \\
W &=& W_q + W_g = 0. \label{work_balance}
\end{eqnarray}
In view of Eqs.~(\ref{work_balance}) and (\ref{work}), one finds the total pressure to be zero, i.e. $P = p_q + p_g = 0$. It is
thus concluded by C. Lorce that the quark gives a positive pressure ($p_q > 0$) and is balanced by the negative pressure from the glue part ($p_g < 0$) to have a stability condition for the nucleon, and hadrons in general.

It has been pointed out by Y. Hatta, A.~Rajan and K.~Tanakain~\cite{Hatta:2018sqd} that the renormalization and trace taking of individual quark and glue EMT do not commute in dimensional regularization, i.e.,
\begin{equation}
\eta_{\mu\nu}(T_{q, g}^{\mu\nu})_R \neq (T_{q,g\, \mu}^{\mu})_R,
\end{equation}
and the perturbative mixing of $\eta^{\mu\nu}T_{q, g}^{\mu\nu}$ has been worked out to two loops~\cite{Hatta:2018sqd} 
and three loops~\cite{Tanaka:2018nae}. Taking this into account, $\bar{C}_{q,g}$ in one loop becomes
\begin{eqnarray}   \label{C_term}
\bar{C}_{q_f}(0, \mu) &=& \frac{1}{4} \left\{   f_f^N + \frac{\alpha_s}{4\pi} \left[ 
\frac{\langle (G^2)_R\rangle}{3 M_N} + \frac{4 C_F}{3} f_f^N \right] - \langle x \rangle_{q_f}\right\}, \\
\bar{C}_{g}(0, \mu) &=& \frac{1}{4} \left\{ \frac{\alpha_s}{4\pi} \left[ 
- \frac{11C_A}{6} \frac{\langle (G^2)_R\rangle}{M_N} + \frac{14 C_F}{3} \sum_f f_f^N \right] - \langle x \rangle_{g}\right\},
\end{eqnarray}
Using $\gamma_{m_0} = 6\, C_F$ and $\alpha_s (2\, \rm{GeV}) = 0.301$ and $\alpha_s (250\, \rm{GeV}) = 0.102$, we obtain
$\bar{C}_q (2\, \rm{GeV}) = -0.161(10) , \bar{C}_q (250\, \rm{GeV}) = -0.122(47)$, which are close to the asymptotic value
of $- 0.146$ for $n_f =3$ in the chiral limit as calculated in Ref.~\cite{Hatta:2018sqd}. Separating out the quark from
the glue contributions in the trace, it introduces a scheme dependence in $\bar{C}(0, \mu)$ as is in the
case of decomposition with the Hamiltonian in Eq.~(\ref{4-term_trace}). On the other hand, if one does not
separate out $H_m$ from $H_a$ in the trace, the rest energy from the internal energies 
\begin{equation} \label{U_3_term}
E_0 = U_q (\mu)+ U_g (\mu)= \big[\frac{3}{4}\langle x\rangle_q (\mu)+ \frac{3}{4}\langle x\rangle_g (\mu) \big] M + \frac{1}{4} M,
\end{equation}
has the same decomposition as the rest energy in Eq.~(\ref{3-term_energy}) from the Hamiltonian.

\section{Trace Anomaly, vacuum energy, and Cosmological Constant}  \label{cosmological_constant}
 
   Throughout the discussion of the nucleon mass and rest energy decompositions, the role of the trace anomaly
remains mysterious. Besides giving rise to scale breaking, it is not clear what physical role it plays as far as the
hadron mass and structure are concerned. In the mass decomposition, the hadron mass is entirely due to the trace anomaly at
the chiral limit in Eq.~(\ref{Mass_total}). This does not reveal any dynamical information of the trace anomaly.
On the other hand, as seen from the decomposition of the rest energy from the Hamiltonian in Eqs.~(\ref{3-term_energy}) and (\ref{4-term_trace}) and the gravitational form factors in Eq.~(\ref{U_3_term}), there are other terms besides the trace
anomaly that may make them more susceptible to divulging the dynamical origin and function of the trace anomaly.

As we shall see, it would be helpful to describe the nucleon in thermodynamic principles. For that, we need to first establish that the nucleon can be treated as a statistical system. To this end, we shall look at the nucleon from the path-integral formulation of QCD in the Euclidean space. QCD in Euclidean path-integral formulation has the same form as that of the classical statistical mechanics. The vacuum-to-vacuum transition in the quantum field theory is termed the grand canonical partition function
\begin{equation} \label{grand}
Z_{GC}(V,T,\mu)= \int \mathcal{D}U \mathcal{D}\bar{\psi} \mathcal{D}\psi\, e^{-S_G(U) - S_F(U,\bar{\psi},\psi,\mu)},
\end{equation}
 where $S_G$ and $S_F$ are the gauge and fermion actions. $T$ is the temperature and $\mu$ is the chemical potential. There
 are infinite degrees of freedom in the partition function. Extensive lattice calculations have been carried out
 to study the QCD phase diagram in finite temperature and chemical potential (for a review of the status of the lattice
 calculation, see for example Ref.~\cite{Philipsen:2019rjq}). The canonical ensemble approach has been
 formulated~\cite{Liu:2002qr,Alexandru:2005ix} to address the problem with a definite baryon number. This entails the projection of the fermion determinant from $Z_{GC}(V,T, \mu)$ with imaginary $\mu$. Since the center $Z_3$ symmetry is preserved in
 the canonical approach, the projected quark numbers are multiples of 3, thus the canonical partition $Z_C(V,T,n_B)$
 is a function of the baryon number $n_B$.  The chemical potential at fixed baryon number $n_B$ is 
 \begin{equation}
 \mu (n_B) = - \frac{1}{\beta} \ln \frac{Z_C (n_B + 1)}{Z_C(n_B)}= \frac{F_{n_B + 1} - F_{n_B}}{(n_B+1) - n_B},
 \end{equation}   
where $F_{n_B}$ is the free energy. For one
nucleon, $\mu\, (n_B = 1)$ is the nucleon mass, i.e. $\mu\, (n_B = 1) = M$ when $T \rightarrow 0$. This is the same
as calculating the nucleon mass from the two-point nucleon correlator with $t \rightarrow \infty$ in the grand canonical ensemble
in Eq.~(\ref{grand}) at $\mu =0$. The quark matrix element of the nucleon can be similarly obtained with projected quark 
propagators~\cite{Alexandru:2005ix}. A first-order phase transition at finite density and temperature is clearly seen with 
the Maxwell construction and the critical point is determined~\cite{Li:2011ee}. However, this was done on a small
$4^4$ lattice. Once the volume is increased, the sign problem sets in abruptly that impedes the lattice calculation.
     
     Given the above formalism, we can discuss the nucleon thermodynamic properties from the canonical ensemble.
We first notice that the $\bar{C}$ terms are associated with the metric $\eta^{\mu\nu}$ in the gravitational form factor
in Eq.~(\ref{EMT_FF}) which are like the cosmological constant, except that the sum $\bar{C}_q + \bar{C}_g = 0$.
However, we need to examine Eq.~(\ref{bar-C}) further and point out that the origin of $f_f^N$ and $f_a^N$ 
is different from that of  $\langle x\rangle_q$ and $\langle x\rangle_g$ in $\bar{C}_q$ and
$\bar{C}_g$. Instead of decomposing the EMT in terms of quark and glue contributions, it turns out to be more 
fruitful to examine the separation of the traceless and trace contributions in Eq.~(\ref{trace_separation}). We first note that there is a glue condensate in the vacuum with a negative energy density (i.e. $\langle 0|G^2|0\rangle < 0$) so is there a negative quark condensate in the vacuum (i.e. $\langle0| \bar{\psi}\psi|0\rangle <0$). The former is from the
conformal symmetry breaking and the latter from the chiral symmetry breaking. One can picture the hadrons
as bubbles in the sea of the condensates. The hadron gas can go through a first-order phase transition to become
the quark-gluon plasma at finite temperature and chemical potential, much like the bubbles turning into steam
during the first-order phase transition between water and steam. As such, the trace anomaly can be considered the vacuum energy of the hadron due to the fact that the bubble with a finite proper volume is created from the surrounding condensate with
negative energy.
In other words, 
\begin{equation}
E_{vac} \equiv \langle H_a\rangle =  \epsilon_{vac} V, 
\end{equation} 
where $\epsilon_{vac} = - \langle 0|H_a|0\rangle$ is the vacuum energy density and $V$ is the proper three-volume of the hadron. Similarly, the disconnected insertion part of the sigma terms arise from the quark condensates in the vacuum.
Since $\langle H_m\rangle$ is small compared to $\langle H_a\rangle$---especially the connected insertion part, in the nucleon at the hadronic scale as shown in Table~\ref{tab:mass} and Fig.~\ref{mass_3f}, we shall ignore it. As the volume changes, we have, from the first law of thermodynamics
\begin{equation}
d E_{vac} = - P_{vac}\, dV 
\end{equation}
 at zero temperature where $d Q = T dS =0$. Here the pressure $P_{vac}$ is
\begin{equation}
P_{vac} = - \epsilon_{vac} < 0,
\end{equation}
%
which is negative. On the other hand, the
other energies $\langle H_E (\mu)\rangle + \langle H_g (\mu)\rangle $ in Eq.~(\ref{sum_rule_4_term}) 
or $\langle x\rangle_{q,g}$ in Eqs.~(\ref{f_qf}) and (\ref{f_g}) are ascribed to the quark kinetic and potential energy and the glue
field energy. Their energy densities should go down if the hadron is allowed to expand so that their energy densities are
like those in the cosmological models where the matter density falls off like $1/a^3$ and radiation density falls off
like $1/a^4$~\cite{Frieman:2008sn}, where $a$ is a scale factor for the radius of the Universe.. This will give a positive pressure to balance that from the $\langle H_a\rangle$ so that the nucleon and other ground state hadrons in QCD are stable. We can parametrize the rest energy (or internal energy) of the nucleon with two terms to illustrate the situation.
\begin{equation}  \label{E_V}
E_0 = \epsilon_{vac} V + \epsilon_{mat} V^p,
\end{equation}
where the first term is from the anomaly that is proportional to the volume and the second term represents the quark kinetic
and potential energies and the glue field energy as in Eqs.~(\ref{sum_rule_4_term}) and (\ref{3-term_energy}). It is proportional to $V^p$ with $ p < 0$. Taking the
derivative with respect to $V$ gives the pressure which should be canceled between the two contributions
\begin{equation}   \label{p-volume}
P_{vac} + P_{k} = - \frac{d E_0}{dV}  = - \epsilon_{vac} -  p\, \epsilon_{mat} V^{p-1} = 0.,
\end{equation}
Given that $E_0 = 4 E_S$ from the Virial theorem consideration~\cite{Ji:2021mtz}, where $E_S = \frac{1}{4} (\langle H_a\rangle + \langle H_m\rangle)$ is the scalar part of the energy, one obtains $p = - \frac{1}{3}$. This is the consequence of the Virial theorem for the scalar and tensor energies. One notices that Eq.~(\ref{E_V}) is exactly like the MIT bag model~\cite{Chodos:1974je} where $\epsilon_{vac}$ is the bag constant $B$ which provides the confinement and $\epsilon_{mat} V^{-1/3}$ corresponds to the quark and gluon normal modes in the bag cavity which are proportional to $1/R$ where $R$ is the bag radius~\cite{Chodos:1974pn}. 

More importantly, we notice that the pressure-volume equation in Eqs.~(\ref{C-pressure}), (\ref{bar-C}) and (\ref{work_balance})  can be written as
\begin{equation}  \label{P-total}
 P_{\rm{total}}=  - \frac{d\,E_0}{d\,V} =  - \frac{E_S}{V} + \frac{1}{3} \frac{E_T}{V}  = 0,
\end{equation}
where $E_T$ and $E_S$ are the tensor (traceless) and scalar (trace) parts of the energy $\langle T^{00}\rangle$ in
Eq.~(\ref{trace_separation}), i.e.
\begin{equation}
E_0 = E_T + E_S,
\end{equation}
where
\begin{eqnarray}
E_T &=& \langle H_{q_f} (\mu)\rangle +   \langle H_g (\mu) \rangle = \frac{3}{4} \left[ \sum_f \langle x\rangle_f (\mu) +
  \langle x\rangle_g (\mu) \right] M, \\
  E_S &=& \frac{1}{4} [\langle H_m \rangle + \langle H_a\rangle].
\end{eqnarray}
Notice that the $ \frac{1}{3}$ factor in Eq.~(\ref{P-total}) comes from the fact that $E_T = 3 E_S$ and it coincides with
the $p = - \frac{1}{3}$ from the volume dependence in Eq.~(\ref{p-volume}). Thus the two pressure equations,
Eqs.~(\ref{P-total}) and (\ref{p-volume}), are consistent with each other so that one can conclude that the scalar energy density $\frac{E_S}{V} = \epsilon_{vac}$ is a constant.  On the other hand, the tensor energy density is not a constant and $E_T$ is proportional to $V^{-1/3}$. Since $E_S$ appears in the metric term in the EMT, it yields a constant restoring pressure to balance that from the tensor energy $E_T$,  leading to the confinement of hadrons. 

This is analogous to the cosmological constant  $\Lambda$ that Einstein introduced to his equation in general relativity~\cite{Einstein:1917ce} 
\begin{equation} \label{Einstein-eq}
R_{\mu\nu} - \frac{1}{2} R\, g_{\mu\nu} + \Lambda\, g_{\mu\nu} = 8\pi G\, T_{\mu\nu}
\end{equation}
where $R_{\mu\nu}$ is the Ricci curvature tensor and $R$ is the scalar curvature. $G$ is the Newton's constant and
the source $T_{\mu\nu}$ is the energy-momentum tensor. The positive constant $\Lambda$ is introduced to the $g_{\mu\nu}$ term so that it balances the gravitational pull from a static uniform matter density $\rho$. Einstein
found the solution of $\Lambda$ to be~\cite{Einstein:1917ce,ORaifeartaigh:2017uct}
\begin{equation}
\Lambda = 4\pi G \rho.
\end{equation}
Later on, when the Universe was found to expand by Hubble, Einstein considered the introduction of the cosmological constant the biggest blunder of his life. The confinement mechanism with the hadrons is somewhat different from that of the
 the cosmological constant $\Lambda$ in Eq.~(\ref{Einstein-eq}) in that the pressure from the energy density of $E_S$ is negative so that it balances the positive pressure exerted by the quark and glue kinetic energy from $E_T$. There have been suggestions that the glue 
part of the trace anomaly is the vacuum energy which is responsible for the confinement of 
hadrons~\cite{Shifman:1978by,Shuryak:1978yk,Teryaev:2016edw,Ji:1995sv,Ji:2021pys}. We have proved from the stress-pressure equation in Eq.~(\ref{P-total}) that the energy density of $E_S$ is indeed a constant that leads to a constant restoring pressure that is responsible for the confinement of hadrons. 

Heavy quarkonium is an analogous but more familiar case.
Since the lattice simulation reveals an area law of the Wilson loop that displays a linearly confining potential between the
infinitely heavy quark-antiquark pair~\cite{Bali:1997am} and the flux tube is formed along the separated color sources~\cite{Baker:2018mhw},
they are consistent with a picture of a constant vacuum energy density for $\langle H_a\rangle$. When the flux tube has a fixed
cross section $A$, the static potential between the heavy quark-antiqurak pair is
\begin{equation}  \label{linear_potential}
V(r) = \epsilon_{vac}\, A \,r = \sigma\, r
\end{equation} 
 which is the origin of the linearly rising potential, where $\sigma$ is the string tension. We can estimate the string tension 
 $\sigma$ from the $G^2$ term of the trace anomaly in the charmonium. The potential between infinitely heavy quark-antiquark pair can be calculated from the rectangular Wilson loop $W_L(r,T)$ which is a plaquette with time $T$ and spatial extent $r$. 
\begin{equation}
V(r) \equiv \lim_{T \rightarrow \infty} \frac{\log \langle W_L(r,T)\rangle}{T},
\end{equation}

It is shown that under the renormalization group consideration, the potential and its derivative is related
to the $G^2$ term of the trace anomaly in the quenched approximation~\cite{Dosch:1995fz,Rothe:1995hu}
\begin{equation}  \label{plaquette}
V(r) + r \frac{\partial V(r)}{\partial r}  =  \frac{\langle \frac{\beta}{2g} (\int d^3\vec{x}\, G^2)\, W_L(r,T) 
\rangle}{\langle W_L(r,T)\rangle}.
\end{equation}
For the charmonium, we can check this relation to see how well it holds. In this case, the corresponding
matrix element for the right-hand side of Eq.~(\ref{plaquette}) is
\begin{equation}
\langle H_{\beta}\rangle_{\bar{c}c} =  \frac{\langle \bar{c}c| \frac{\beta}{2g} \int d^3\vec{x}\, G^2 |\bar{c}c\rangle}
{\langle \bar{c}c|\bar{c}c\rangle}.
\end{equation}
For the linear potential in Eq.~(\ref{linear_potential}), this implies
\begin{equation}  \label{sigma}
2  \sigma \langle r \rangle = \langle H_{\beta}\rangle_{\bar{c}c},
\end{equation}
for the potential energy. $\langle H_{\beta}\rangle_{\bar{c}c}$ can be obtained from the lattice calculation of $\langle H_m\rangle_{\bar{c}c}$,
\begin{equation}   \label{H_beta}
\langle H_{\beta}\rangle_{\bar{c}c} = M_{\bar{c}c} - (1+ \gamma_m) \langle H_m\rangle_{\bar{c}c}.
\end{equation}
Taking $M_{c\bar{c}}$ to be the spin average of the $J/\Psi$ and $\eta_c$ masses,  $M_{c\bar{c}} = 3069$ MeV.
The sigma term in the charmonium in a recent lattice calculation of the charmonium structure~\cite{Sun:2020pda} has been obtained to be $\langle H_m\rangle = 2166(1)$ MeV.  At the scale of the lattice spacing $a^{-1} = 1.785$ GeV used in the
the lattice calculation~\cite{He:2021bof,Sun:2020pda}, $\gamma_m (\mu = 0.1785\, {\rm GeV})= 0.325$. This is consistent with a fit of the charmonium spectrum with the Cornell potential that determined the effective $\alpha_s = 0.491(80)$~\cite{Mateu:2018zym}, and this implies that the leading order $\gamma_m = 6 \,C_F \frac{\alpha_s}{4\pi} = 0.31$. With $\gamma_m = 0.325$, one can deduce from Eq.~(\ref{H_beta}) that $\langle H_{\beta}\rangle = 199$ MeV. 

To estimate $\langle r\rangle$ in Eq.~(\ref{sigma}), we shall use the ratio $\frac{\langle r\rangle}{\langle r^2\rangle^{1/2}}$. Since the asymptotic behavior of the $1S$ wave function of a linear potential in the Schr\"{o}dinger equation is between that
of the harmonic oscillator and the Coulomb potential, we shall use the average of the above ratios from the $r^2$ and
$ - 1/r$ potentials, i.e. 
$\frac{\langle r\rangle}{\langle r^2\rangle^{1/2}} = \frac{1}{2} [\frac{\langle r\rangle}{\langle r^2\rangle^{1/2}}|_{\rm{H.O.}} + \frac{\langle r\rangle}{\langle r^2\rangle^{1/2}}|_{\rm{Coul}}]= 0.61$. In a potential model to fit the charmonioum spectrum which includes both the linear and Coulomb potentials and the spin-spin and spin-orbit interactions, $\langle r^2\rangle^{1/2} = 0.21$fm ~\cite{Liu:1979de}. This gives $\langle r\rangle \sim 0.13$ fm. From Eq.~(\ref{sigma}) and $\langle H_{\beta}\rangle$ from
the lattice calculation in Eq.~(\ref{H_beta}), we obtain the string tension $\sigma = 0.153\, {\rm GeV^2}$.
This is in very good agreement with that from the recent Cornell potential analysis of the charmonium spectrum which gives the fitted string tension $\sigma = 0.164(11)\, {\rm GeV^2}$. Despite this close agreement, we should caution that Eq.~(\ref{plaquette}) is for infinitely heavy quarks in a pure gauge theory, while the lattice calculation~\cite{He:2021bof,Sun:2020pda} is carried out with realistic finite charm quark mass on (2+1)-flavor dynamical fermion configurations.

An effective low-energy theory in curved spacetime with broken scale invariance has been formulated which can lead to 
confinement~\cite{Kharzeev:2004ct,Kharzeev:2008br}.  The trace anomaly (quantum anomalous energy) contribution
to the nucleon mass has been considered in a Higgs mechanism~\cite{Ji:2021pys}. Here, we identify
$\frac{1}{4}\langle H_{\beta}\rangle$ as the vacuum energy emerged from the formation of a hadron bubble in the sea of the glue condensate. We should emphasize the fact that the vacuum energy density is a constant, so is its negative restoring pressure which inevitably results in confinement. This is  similar to the case where the constant force from a linear potential confines the heavy quarkonium.

The cosmological constant in general relativity has a  renewed interpretation after the Universe is found to have an
accelerating expansion. The pressure from the energy-momentum tensor also gravitates and contributes to the acceleration of the expanding Universe. The Freidman equation from the Friedmann-Robertson-Walker metric is
\begin{equation}
\frac{\ddot a}{a} = - \frac{4\pi G}{3} (\rho + 3 P),
\end{equation}
where $a$ is a scale factor, $\rho/P$ is the density/pressure from all matter, radiation, and dark energy. The cosmological
constant, which would appear as $\Lambda/3$ on the right-hand side for historical reasons, has been subsumed into the $\rho$
and $p$ as vacuum energy and pressure with $\rho_{vac}= \Lambda/(8\pi G)$ and $p_{vac} = - \Lambda/(8\pi G)$.
When the negative pressure from the dark energy (cosmological constant) overcomes  $\rho$, the Universe expansion
accelerates. In the case of QCD, the negative pressure $P_{vac}$ simply balances out the pressures from the quark kinetic
energy and glue field energy and confines the quarks and gluons. In this sense, the hadron is analogous to the static universe that Einstein had imaged for the cosmological constant to achieve and $E_S/V$ or, more specifically, $\langle H_{\beta}\rangle$
plays the role of the cosmological constant for the hadrons.

It would be interesting to observe the glue part of the trace anomaly $\langle H_{\beta}\rangle $ experimentally~\cite{Meziani:2020oks}. 
It is shown by D. Kharzeev~\cite{Kharzeev:1995ij} that the photoproduction of $J/\Psi$ at threshold would be a place to probe
this. It has also been explored by Y. Hatta and D.L. Yang, using gauge/string duality~\cite{Hatta:2018ina}.
One could also explore the trace anomaly and conformal symmetry on the lattice. Through the study of the Dirac eigenvalue density, there is evidence that there is a phase above the crossover temperature that displays infrared scale 
invariance~\cite{Alexandru:2019gdm}. It would be useful to find out what impact it may have on $\langle H_{\beta}\rangle$. There are efforts to look for conformal window with multiflavor simulations~\cite{DelDebbio:2010zz}. One could ask the same question about $\langle H_{\beta}\rangle$. Also, using it as an indicator, one could calculate it in nuclei to see if the conformal symmetry 
is partially restored. 


\section{Summary}   \label{summary}

    We have considered the decomposition of the proton mass and rest energy in terms of their quark and glue components.
The proton mass from the trace of the EMT has a unique division in terms of the quark sigma terms and the quantum trace anomaly that are scheme and scale independent. Since the mass is not additive in general frames, the decomposition of the proton mass in terms of expectation values can be carried out in the rest frame or a comoving frame. 
We use the lattice results to enumerate each term. The role of the heavy quarks is clarified. 

There are different ways to decompose the rest energy, be it from the Hamiltonian or the gravitational form factors. They are scheme and scale dependent. The simplest and least scheme dependent way is to divide the Hamiltonian in terms of the trace and the traceless parts, which is the same as obtained from the forward gravitational form factors. The traceless part can be separated into quark and glue momentum fractions, measurable from DIS and Drell-Yan experiments. We use the CT18 global fitting of $\langle x\rangle_q$ and $\langle x\rangle_g$ to display this decomposition at 
$\mu = 2$ and 250 GeV. We note that the trace anomaly introduces a scale in QCD and its mass decomposition can be expressed in terms of the sigma terms and the trace anomaly. The proton mass from the trace of the EMT is simply four times the scalar energy, i.e. $M = 4 E_S$.

    One interesting feature is revealed from the decomposition of the gravitational form factors. The forward $\eta^{\mu\nu}$ (metric) term $\bar{C}_{q,g}$ is the normal stress of the EMT which is the negative of the pressure for the system. The total pressure from the trace part of the rest energy ($E_S$) and the traceless part ($E_T$) is zero due to the conservation of the EMT. It shows that
the pressure from $E_S$ is canceled by $-\frac{1}{3}$ of that of $E_T$. This reflects the fact that $E_T = 3 E_S$.
More importantly, expressed in terms of their volume dependence, it discloses that $E_S$ is linearly dependent on
volume which gives a constant restoring pressure to balance the positive pressure from $E_T$ which has a volume dependence
of $V^{-1/3}$. $E_S$ with constant energy density is naturally interpreted as the vacuum energy from the condensate;
in particular, the glue part of the trace anomaly, which dominates $E_S$, is due to the forming of a nucleon bubble in
the sea of the glue condensate. Einstein introduced the cosmological constant in general relativity with an intention to describe 
a static universe. It is actually more applicable to delineate the confinement of hadrons.

   The linear potential between the heavy quark-antiquark pair can be understood as due to a constant vacuum energy
density and a flux tube formation between the heavy quarks. We deduce the string tension from the glue part
of the the trace anomaly $\langle H_{\beta}\rangle_{\bar{c}c}$ in a lattice calculation of the charmonium and found it to be in very good agreement with
that from a Cornell potential that fits the charmonium spectrum. This further supports the notion that the glue part
of the trace anomaly is the hadron cosmological constant which is responsible for hadron confinement.
Studies of the trace anomaly in conformal symmetry restored phase in QCD and in conformal field theories
may shed light on this issue.

\newpage
\section{Acknowledgment}
The author is indebted to X. Ji, C. Lorc\'{e}, Y. Hatta, D. Kharzeev, A. Metz, B. Pasquini, S. Rodini, M. Constantinou, M. Polyakov, P. Schweitzer, and Y.B. Yang for fruitful discussions. He also thanks T.J. Hou for providing the CT18 data and G. Wang for
help with the figures.This work is partially supported by the U.S. DOE Grant No. DE-SC0013065 and No.\ DE-AC05-06OR23177 which is within the framework of the TMD Topical Collaboration.
This research used resources of the Oak Ridge Leadership Computing Facility at the Oak Ridge National Laboratory, which is supported by the Office of Science of the U.S. Department of Energy under Contract No.\ DE-AC05-00OR22725. This work used Stampede time under the Extreme Science and Engineering Discovery Environment (XSEDE), which is supported by National Science Foundation Grant No. ACI-1053575.
We also thank the National Energy Research Scientific Computing Center (NERSC) for providing HPC resources that have contributed to the research results reported within this paper.
We acknowledge the facilities of the USQCD Collaboration used for this research in part, which are funded by the Office of Science of the U.S. Department of Energy.


\end{document}